\documentclass{article}

\usepackage{newlfont}
\usepackage{amsfonts}
\usepackage{amsmath}
\usepackage{graphicx}
\usepackage{setspace}
\usepackage{cases}
\usepackage{stmaryrd}
\usepackage{hyperref}
\usepackage{epstopdf} 
\usepackage{mathrsfs}

\newcommand{\R}{\mathbf{r}}
\newcommand{\N}{\mathbf{n}}
\newcommand{\E}{\mathbf{e}}
\newcommand{\Y}{\mathbf{y}}
\newcommand{\G}{\mathbf{G}}
\newcommand{\h}{\mathbf{H}}
\newcommand{\Gam}{\boldsymbol\Gamma}
\newcommand{\X}{\mathbf{x}}

\newcommand{\F}{\mathbf{F}}

\newcommand{\B}{\mathbf{b}}
\newcommand{\A}{\mathbf{a}}

\newcommand{\pa}{\! \! \Arrownot \, \, \Arrownot \, \, \, \,}

\begin{document}

\title{\textbf{A study of snake-like locomotion  \\ through the analysis of \\
a flexible robot model} }
\author{Giancarlo Cicconofri and Antonio DeSimone \\
SISSA, International School for Advanced Studies
\\
 Via Bonomea 265, 34136 Trieste - Italy
\\
giancarlo.cicconofri$@$sissa.it , desimone$@$sissa.it}

\date{}
\maketitle

\begin{abstract}
We examine the problem of snake-like locomotion by studying a system consisting of a planar inextensible elastic rod with adjustable spontaneous curvature, which provides an internal actuation mechanism that mimics muscular action in a snake. 

Using a Cosserat model, we derive the equations of motion in two special cases: one in which the rod can only move along a prescribed curve, and one in which the rod is constrained to slide longitudinally without slipping laterally, but the path is not fixed a-priori (free-path case). The second setting is inspired by undulatory locomotion of snakes on flat surfaces.

The presence of constraints leads in both cases to non-standard boundary conditions, that allow us to close and solve the equations of motion. The kinematics and dynamics of the system can be recovered from a one-dimensional equation, without any restrictive assumption on the followed trajectory or the actuation. We derive explicit formulas highlighting the role of  spontaneous curvature in providing the driving force (and the steering, in the free-path case) needed for locomotion.

We also provide analytical solutions for a special class of  serpentine motions,  which enable us to discuss the connection between observed trajectories, internal actuation, and forces exchanged with the environment.

\end{abstract}

\newpage

\tableofcontents

\newpage

\section{Introduction}

Snake locomotion has fascinated natural scientists for a long time.
More recently, it has become a topic of great interest as one of the key examples of soft bio-inspired robotics. This is a new and recent paradigm in robotic science \cite{TrRev,Kim}, whereby inspiration is sought from nature to endow robots with new capabilities in terms of dexterity (e.g.,  the manipulation abilities of an elephant trunk or of an octopus arm) and adaptability (e.g., the ability of snakes to handle unexpected interactions with unstructured environments and move successfully on uneven terrains by adapting their gait to ground properties that change from place to place in an unpredictable way). 

The way snakes move has been the subject of seminal works by Gray \cite{Gray,GrayLiss}, see also \cite{Be,McNeil}. In these early studies Gray described the mechanics underlying snake locomotion inside closely fitting channels and on a surface in the presence of external push-points. Subsequently,  muscular activity as well as forces transmitted by snakes to arrays of pegs among which they move have been measured  \cite{Jayne,Mo}. Further early theoretical studies can be found in the Russian literature, see, e.g., \cite{Lav,Kuz,Che1,Che2} and the references quoted therein. More recently, focus has turned to the importance of frictional anisotropy between snakes ventral skin and flat surfaces on which they move, stimulating both experimental and theoretical research \cite{Bau,Ber,HuShe,HuShe2,Mah}. In fact, it is well established that equality of friction coefficients in longitudinal and lateral directions leads to no net forward motion in undulatory locomotion (see, e.g., \cite{HuShe,HuShe2}, and  \cite{Nlink,Morandotti2} for similar results in the closely related problem of undulatory swimming locomotion).

The idea that frictional anisotropy plays a role in snake locomotion was put forward long ago in the engineering literature \cite{Be} and, most notably, by Hirose in his seminal work on robotic snake-like locomotion \cite{Hi}. Hirose was among the first to realize the potential of biological inspiration in designing robots by  studying snake-like locomotors and manipulators \cite{Hi}. Technological advances in this field have led to the development of models  for snake robots crafted with more and more jointed active segments, eventually leading to the use of continuum theories \cite{Chi}. In some more recent contributions \cite{BoySha,BoyPo,Tr,Re,Rucker}, Cosserat models are used for the mechanics of slender flexible robots, described as deformable rods.

Inspired by the literature on snake-like locomotion recalled above, in this paper we study a model system similar to the one used in \cite{Fauci} in the context of undulatory swimming, and consisting of a planar inextensible elastic rod that is able to control its spontaneous curvature. This is the curvature the rod would exhibit in the absence of external forces, which can be non-zero in the presence of internal actuation (see the sketch in Fig.\ref{1}B). Local control of this quantity provides an internal actuation mechanism that can be used to mimic muscular activity in biological undulatory locomotion. Indeed, by varying its spontaneous curvature $\alpha$, the rod generates a distributed internal bending moment $M^{a}$. The two quantities satisfy the simple relation $M^{a}=-EJ \alpha$, where $EJ$ is the bending stiffness of the rod, see \eqref{a01}.  Travelling waves of spontaneous curvature can put the system in motion when the environment exerts constraints or forces that prevent the rod to be deformed everywhere according to its spontaneous curvature.

To show how control of spontaneous curvature in the presence of external constraints leads to locomotion we use a Cosserat model, and derive the equations of motion for two special cases: 
one in which the rod can only move along a prescribed curve (prescribed-path case), and one in which the rod is constrained to slide longitudinally without slipping laterally, but the path is not fixed a-priori (free-path case). 
The first case corresponds to a rod confined in a channel with frictionless walls.
The second case is inspired by the slithering motion of snakes, that interact through anisotropic frictional forces with a flat surface on which they are free to move. Frictional resistance is typically larger in the lateral direction than in the longitudinal one.
Our setting corresponds to the limiting case of infinite ratio between lateral and  longitudinal  friction coefficients, in which longitudinal sliding is allowed while lateral slipping is forbidden.

Our work is closely related to the approach presented in \cite{Mah}, which we extend in at least one major way. In fact, in  \cite{Mah} locomotion of an active rod with no lateral slipping along a free path is considered. The trajectory followed by the rod is an unknown of the problem. The authors impose, however, periodicity of the solution (effectively considering a rod of infinite length) which leads to an incomplete system of equations. The system is then closed  by postulating laws (closure relations, justified by experimental observations) on the lateral forces exerted on the ground surface. The novelty of our approach consists in solving the equations of motion in the case of a system of finite length, with no a-priori assumptions on either the followed path, which can be non-periodic, or on the reactive forces imposing no lateral slipping. These emerge both as part of the solution of the problem, once a history of spontaneous curvatures is assigned.
Closure of the equations is obtained by carefully considering edge-effects, which lead to non-standard boundary conditions. We derive in this way explicit formulas  that enable us  to explore in full generality the connection between observed motion, internal actuation, and lateral forces exchanged with the environment. Moreover, we are able to solve inverse locomotion problems, namely,  given a motion of the system that we want to observe, find an internal actuation that produces it.

Our main results are the following. We formulate direct and inverse locomotion problems (direct: find the motion produced by a given actuation history; inverse: find the actuation history required to produce a given motion),  and show existence and uniqueness of the solution of direct problems, non-uniqueness for the inverse ones. In the prescribed-path case, we reduce the dynamics of the system to a single ordinary differential equation for the tail end coordinate (the only degree of freedom for an inextensible rod forced to slide along a given curve). This equation reveals clearly the mechanism by which a flexible rod can actively propel itself inside a channel, whenever the channel exhibits a variation of curvature along its track, and provides a quantitative framework to revisit some of the classical findings on snake motility by Gray. 

In the free-path case, we are again able to close the equation of motion and reduce the dynamics of the system to a single equation, this time an integro-differential equation for the tail end coordinate. A particularly interesting outcome of our analysis  is the emergence of an asymmetry in the mechanical boundary conditions at the (leading) head and the (trailing) tail. This is not only a mathematical subtlety, but it is also deeply grounded in the physics of the problem. While the tail follows the path traced by the preceding interior points, the head is free to veer laterally, `creating' the path as the motion progresses. We show that the curvature of this newly created path is set by the time history of spontaneous curvatures at the leading head. Recognising this steering role of the spontaneous curvature leads to a procedure to generate solutions for the free-path case from those of the prescribed-path case,  based on  modifying them near the leading head, in order to account f
 or steering. Again, we provide explicit formulas to calculate the lateral forces transmitted to the ground surface.

The rest of the paper is organized as follows.
In Section 2 we present our mathematical model of flexible robot as an active rod, and formulate direct and inverse locomotion problems. In Section 3 we derive the governing equations and the appropriate boundary conditions for motion inside a channel with frictionless walls (prescribed-path), solve them in some simple geometries, and discuss the physical implications of our results. In Section 4 we derive the governing equations and 
corresponding boundary conditions for the motion of an active rod sliding longitudinally without slipping laterally on a flat surface (free-path) and propose a class of analytical serpentine solutions. Possible connections of our  results with observations made in the context of biological snake locomotion are briefly summarised  in the Discussion section, while the existence and uniqueness of the solution  of the equations of motion for the free-path case is proved in the Appendix.

\section{The flexible robot model}
\begin{figure}[h]
\includegraphics[width=1.00\textwidth]{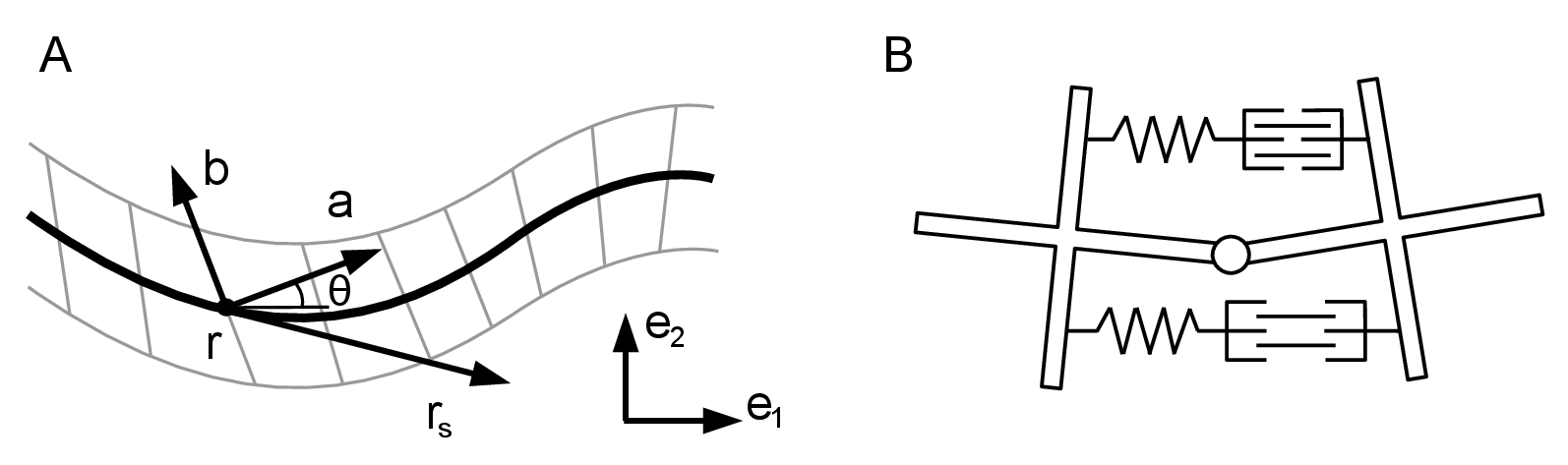}
\caption{A) Variables describing a  Cosserat rod configuration. The cross sections of the continuous rod are depicted through  grey segments transversal to $\R$ (black curve). B) Schematic model for the constitutive elements of the robot structure, illustrating a mechanism to produce non-zero curvature in the absence of external forces.}
\label{1}
\end{figure}

We consider a model consisting of a (long) chain of cross shaped elements (Fig.\ref{1}B) linked together by ideal joints connected by deformable springs. We assume that each spring is able to actively change its rest length (the length at which the tension in the spring is zero). Following \cite{BoyPo,BoySha,Tr,Re,Rucker}  we model this system through a continuous description based on the planar Cosserat rod theory. 

A configuration of a Cosserat rod of reference length $L$ on the plane is defined by a pair of vector-valued functions 
\begin{equation}
\left[0 , L \right] \times \left[0,\infty\right) \, \ni   (s,t) \mapsto \R (s,t) \, , \, \B (s,t)  
\label{conf}
\end{equation}
where $\B$ is a unit vector. The curve $\R$ describes the midline of the rod, while $\B$ characterizes the orientation of its deformed cross sections (see Fig.\ref{1}A). As in \cite{Ant}, we introduce also the unit vector $\A := - \E_{3}\times \B$, where $\E_{3}$ is the unit vector normal to the plane. We then define the  \emph{strain} variables $\nu$ and $\eta$ through the following decomposition along the moving orthonormal frame $\left\{\A ,\B \right\}$
\begin{displaymath}
	\R_{s} = \nu \A + \eta \B
\end{displaymath}
where the subscript $s$ is used  to denote the partial derivative with respect to the space variable. The function $\nu=\nu(s,t)$ describes the \emph{stretch}, while $\eta=\eta(s,t)$ defines the \emph{shear} strain. Finally the \emph{bending} strain $\mu := \theta_{s}$ is obtained through the scalar valued function $\theta(s,t)$ defined by
\begin{displaymath}
	\A (s,t) = \cos \theta (s,t)\E_{1} + \sin \theta (s,t) \E_{2} 
\end{displaymath} 
where  $\left\{\E_{1},\E_{2}\right\}$ is a fixed basis in the plane containing the rod. We consider our system as being made of an infinite number of elements like the ones in Fig.\ref{1}B, each of them being of infinitesimal length, and assembled along the central curve $\R$ of the rod. Since we assume them to be rigid, we impose the constraints that the rod is  inextensible and unshearable:
\begin{equation}
	\nu(s,t) = 1 \quad \textrm{and} \quad \eta(s,t) =0 \, . \label{EB}
\end{equation}
The ability of the robot to modify the equilibrium length of each of the connecting springs can be naturally modelled macroscopically by considering an elastic rod which can actively vary its spontaneous curvature, namely, the curvature the rod would exhibit in the absence of external loads. This is similar to what is done \cite{Fauci} in the context of swimming motility. We model this by introducing the elastic potential density
\begin{equation}
	\mathcal{U}(\mu,s,t) = \frac{EJ}{2}\big(\mu - \alpha(s,t) \big)^{2} \label{ela}
\end{equation}
where $EJ$ is the bending stiffness of the rod. Notice that if (\ref{EB}) hold, then $\R_{s}$ always coincides with the unit vector $\A$, and the bending strain $\mu(s,t)$ is equal to the curvature of the rod at the point $\R(s,t)$. Therefore, the function $\alpha$ in (\ref{ela}) can be viewed as a varying spontaneous curvature, which we assume to be freely controllable in order to set the robot in motion. 
The bending moment resulting from \eqref{ela} is 
\begin{equation}\label{a01}
	M = EJ\big(\mu - \alpha \big) =EJ\theta_{s} + M^{a}
\end{equation}
and can be seen as the sum of a passive elastic term $EJ\theta_{s}$ and of an active one $M^{a}:=-EJ\alpha$ which can be varied at will by suitably tuning $\alpha$. An active moment originating from muscular contraction  is used in the model of snake locomotion in \cite{Mah}.

Along with the elastic potential we define  the kinetic energy density
\begin{displaymath}
	\mathcal{T}(\R_{t},\theta_{t}) = \frac{\rho A }{2} \, \R_{t} \cdot \R_{t} + \frac{\rho J }{2} \, \theta_{t}^{2}
\end{displaymath}
where the subscript $t$ denotes the partial derivative with respect to time, $\rho A$ is the linear mass density and $\rho J$ the linear moment of inertia. Finally, the Lagrangian density $\mathcal{L}$ of the system reads
\begin{equation}
	\mathcal{L} = \mathcal{T} - \mathcal{U} -  N(\nu -1) - H \eta 
	\label{lag}
\end{equation}
where $N=N(s,t)$ and $H=H(s,t)$ are the reactive internal forces (axial tension and shear force, respectively) enforcing constraints (\ref{EB}).

In the following sections we will consider two types of locomotion problems arising form the interaction of prescribed spontaneous curvature and external constraints. The direct one can be formulated as follows: given a time history of spontaneous curvatures $\alpha(s,t)$, together with initial and boundary conditions, find the motion $\R(s,t)$ of the rod and the forces it exchanges with the environment. In the inverse one, the motion is prescribed, and we want to find a history $\alpha(s,t)$ that produces it, together with the corresponding forces. We will consider two types of external constraints and see that, in both cases, the direct problem has unique solution while, for the inverse one, the solution is not unique.
For studies of swimming locomotion problems conducted in a  similar spirit, we refer the reader to \cite{Morandotti2,Fauci,Alouges08,Alouges13,Morandotti1}.

\section{The case of prescribed path: \\ sliding inside a channel}

The first problem we consider is motion along a prescribed path.
We place our robot model inside a curved channel fitting exactly its body, and we assume that there are no friction forces exerted by the walls of the channel. We model such a setting by imposing the external (holonomic) constraint 
\begin{equation}
\R \in \textrm{Graph} \left\{\Gam\right\} \quad \textrm{or} \quad \phi_{\Gamma} (\R)=0
\label{hol}
\end{equation}
where the equation $\phi_{\Gamma} =0$ defines (we assume, globally) the curve $\Gam$ which we interpret as the central line of the channel. There is no loss of generality in assuming $|\nabla \phi_{\Gamma}|=1$.

\subsection{Derivation of the equations of motion}

We derive the equations of motion through Hamilton's Principle, adding to (\ref{lag}) an external reactive potential $-f\phi_{\Gamma} \! (\R)$, where $f=f(s,t)$ is the Lagrange multiplier enforcing (\ref{hol}). A solution $(\R,\theta)$ must satisfy
\begin{equation}
\delta   \int_{t_{1}}^{t_{2}}  \int_{0}^{L} \mathcal{L} - f\phi_{\Gamma}  (\R) \, \, ds dt = 0
\label{H}
\end{equation}
for every variations $\delta \R$ and $\delta \theta$ defined on $\left[0,L\right]\times\left[t_{1},t_{2}\right]$ and vanishing at its boundary. If we define $\N := N\A + H\B$, the Euler-Lagrange equations we obtain from (\ref{H}) are
\begin{displaymath}
\N_{s}  - f \, \nabla \phi_{\Gamma}(\R) = \rho A \, \R_{tt} \quad , \quad  M_{s} \, \E_{3}  + \R_{s} \times \N = \rho J \theta_{tt} \E_{3}
\end{displaymath}
where  the bending moment $M$ is defined in \eqref{a01}. These are the classical dynamical equations for a planar Cosserat rod (see e.g. \cite{Ant}) with an external force given, in our case, by the transversal reaction imposing constraint \eqref{hol}. We can suppose that our active rod is in frictional contact with the ground. The presence of a longitudinal frictional force per unit length 
\begin{equation}\label{a02}
	\F^{\pa}   = - \gamma^{\pa} \frac{\R_{s} }{| \, \R_{s} \,|} \textrm{Sgn}(\R_{t} \cdot \R_{s}) \, ,
\end{equation}
is handled  by simply adding $\F^{\pa}$, where $\textrm{Sgn}$ denotes the sign function,  to the left hand side of the first equation.

To close the equations of motion we use the  Principle of Mechanical Boundary Conditions (PoMBC) \cite{LiSi}. We define \emph{generalized} edge loads acting on the system by considering the rate at which work is expended at the edges in virtual motions compatible with the constraints, and assume that all generalized edge loads acting on the system are explicitly prescribed. 

In view of (\ref{hol}), we have that
\begin{equation}
	\R(0,t)=\Gam(s_{0}(t)) \: \: , \: \: \R(L,t)=\Gam(s_{L}(t)) \: \: , \: \: \theta(0,t)=\Theta(s_{0}(t)) \: \: , \: \: \theta(L,t)= \Theta(s_{L}(t))  \label{init}
\end{equation}
where $s_{0}$ and $s_{L}$ are the curvilinear coordinates relative to $\Gam$ of the two ends of the rod, which we call \emph{generalized edge coordinates},  and $\Theta$ is the angle between the tangent vector to $\Gam$ and $\E_1$, so that
\begin{equation}
	\Gam(\xi) = \Gam(\xi_{0}) + \int_{\xi_{0}}^{\xi} \cos\Theta(\lambda)\E_{1} + \sin\Theta(\lambda)\E_{2} \, \, d \lambda \, .\label{Theta}
\end{equation}
Now, following the PoMBC, we write the work rate $P_{\textrm{edge}}$ of the edge loads as
\begin{equation}
P_{\textrm{edge}} = \N(s,t) \cdot \R_{t}(s,t)\, \Big|_{s=0}^{s=L} + M(s,t) \theta_{t}(s,t) \, \Big|_{s=0}^{s=L} \, .
\label{row}
\end{equation}
Using (\ref{init}) to derive the expressions for $\R_{t}$ and $\theta_{t}$ at $s=0,L$ we obtain
\begin{multline*}
\quad \quad \quad \quad  P_{\textrm{edge}} = \dot{s}_{L}(t)\Big(\N(L,t) \cdot \Gam_{s}(s_{L}(t))+ M(L,t)k(s_{L}(t))\Big) \\ 
    - \dot{s}_{0}(t)\Big(\N(0,t) \cdot \Gam_{s}(s_{0}(t))  + M(0,t)k(s_{0}(t))\Big) \quad \quad \quad \quad
\end{multline*}
where we used a ``dot'' to denote the time derivative of the generalized coordinates, and $k$ is the curvature of $\Gam$. The coefficients multiplying the \emph{generalized velocities} $\dot{s}_{0}(t)$ and $\dot{s}_{L}(t)$ are the \emph{generalized edge loads} which, by the PoMBC, have to be prescribed. Since we suppose that no external edge forces are doing work on the system at either of the two ends, we enforce the condition $P_{\textrm{edge}}=0$ by setting such loads equal to zero.

Finally, conditions (\ref{EB}) and (\ref{hol}) must be added to the equations of the system. Since the active rod is assumed to be inextensible and unshearable, and its backbone curve $\R$ is forced inside the graph of $\Gam$, the constrained system can be described with only one degree of freedom, namely,  the curvilinear coordinate relative to $\Gam$ of the first end of the robot model. Thus, 
\begin{equation}
	\R(s,t)=\Gam(s_{0}(t) + s) \quad , \quad \theta(s,t)=\Theta(s_{0}(t)+s) \label{RandTheta}
\end{equation}
and substituting these expressions in the equations of motion we obtain, accounting also for longitudinal friction,
\begin{gather}
 N_{s} - k H  - \gamma^{\pa} \textrm{Sgn}(\dot{s}_{0}(t)) = \rho A \, \ddot{s}_{0}(t) \label{eqc1} \\
 k N + H_{s} -  f  =  \rho A \, k\dot{s}_{0}(t)^{2} \label{eqc2} \\
 EJ(k_{s} - \alpha_{s})  + H  =  \rho J (k \ddot{s}_{0}(t) + k_{s} \dot{s}_{0}(t)^{2}  ) \label{eqc3}
\end{gather}
where  $k=k(s_{0}(t)+s)$. As for the boundary conditions, they now read
\begin{equation}
\begin{aligned}
 & N(0,t) + EJ\big(k(s_{0}(t)) - \alpha(0,t)\big)k(s_{0}(t))  = 0  \, ,   \\
 & N(L,t) + EJ\big(k(s_{0}(t)+L) - \alpha(L,t)\big)k(s_{0}(t)+L)  = 0\,.
\end{aligned}\label{BCc}
\end{equation}

Summarizing, in order to solve the (direct) locomotion problem stated at the end of Section 2, we need to find the unknown functions $N(s,t)$, $H(s,t)$, $f(s,t)$ and $s_0(t)$. The equations we have for this purpose are the three equations of motion \eqref{eqc1}-\eqref{eqc2}-\eqref{eqc3}, and the two boundary conditions \eqref{BCc}.
We'll see that, by integrating  \eqref{eqc1},  a first order ordinary differential equation (ODE) in the space variable $s$, we can derive one additional ODE (in the time variable) containing only the unknown $s_0(t)$, which completely determines the motion of the system. This ODE is given below as equation \eqref{s0channel}, or \eqref{formula} in a simplified version.
Once $s_0$ is known, we can use \eqref{eqc3}, \eqref{eqc1} and \eqref{eqc2}, together with the  boundary condition \eqref{BCc} holding at $s=0$, to determine $H$, $N$, and $f$ respectively.

We  show now how to obtain the ODE for $s_{0}(t)$. If we substitute in \eqref{eqc1} the expression of $H$ given by \eqref{eqc3} then, integrating  on the space variable, we have
\begin{align*}
m \ddot{s}_{0} \: & = \: N\,\Big|_{0}^{L} \! + EJ   \int_{0}^{L}  (k_{s} - \alpha_{s})k \, ds - \gamma^{\pa} \textrm{Sgn}(\dot{s}_{0}) L  - \rho J \, R - \rho J \, Q\ddot{s}_{0}\\
               \: & = \: N\,\Big|_{0}^{L} \!  + EJ(k - \alpha)k\Big|_{0}^{L} \!   - EJ   \int_{0}^{L}  (k - \alpha)k_{s} \, ds - \gamma^{\pa} \textrm{Sgn}(\dot{s}_{0}) L - \rho J \, R- \rho J \, Q\ddot{s}_{0}
\end{align*} 
where  $m=\int_0^L \rho A ds$ is the total mass of the rod, 
\begin{equation}
\begin{aligned}
	R(\dot{s}_{0}(t),s_{0}(t)) := & \frac{\dot{s}_{0}(t)^{2}}{2} \left(k^{2}(s_{0}(t)+L) - k^{2}(s_{0}(t)) \right)  \, , \\
	\textrm{and} \: \: Q(s_{0}(t)) := &\int_{0}^{L} \! \! k^{2}(s_{0}(t)+s)ds \, .
	\end{aligned}
\label{RandQ}
\end{equation}
If we now apply (\ref{BCc}) we obtain the equation
\begin{equation}
\begin{aligned}
\big(m + \rho J\, Q(s_{0}(t))\big) \, \ddot{s}_{0}(t) \: =  & \: \frac{EJ}{2}\left( k^{2}(s_{0}(t)) - k^{2}(s_{0}(t) + L) \right) - \gamma^{\pa} \textrm{Sgn}(\dot{s}_{0}(t)) L  \\
                                                            & \: - \rho J \, R(\dot{s}_{0}(t),s_{0}(t))  + EJ  \int_{0}^{L}    \alpha(s,t) \, k_{s}(s_{0}(t) + s) \, ds \label{s0channel}
\end{aligned}
\end{equation}
which, complemented  with initial position and velocity, defines $s_{0}$ uniquely. The shear force $H$ is now uniquely defined by \eqref{eqc3}, while
\begin{equation*}
\begin{aligned}
N(s,t) \: & =  \int_{0}^{s} \! \! \left\{\rho A \, \ddot{s}_{0}(t) + \gamma^{\pa} \textrm{Sgn}(\dot{s}_{0}(t))  +  k(s_{0}(t)+\lambda) H(\lambda,t)\right\} d\lambda \\
&  \quad \quad \quad  - \: EJ\big(k(s_{0}(t)) - \alpha(0,t)\big)k(s_{0}(t)) \,  . 
	\end{aligned}
\end{equation*} 
Using all the expressions above we can recover $f$ from \eqref{eqc2}.

Let us now suppose that our active rod is stiff and slender enough, so that  $E J , \rho A \gg \rho J$. We can then neglect the terms containing $\rho J$ in \eqref{s0channel}, obtaining the simplified equation
\begin{equation}
\begin{aligned}
m\ddot{s}_{0}(t) \: & = \frac{EJ}{2}\left( k^{2}(s_{0}(t)) - k^{2} (s_{0}(t) + L) \right) - \gamma^{\pa} \textrm{Sgn}(\dot{s}_{0}(t)) L \\
 &  \quad \quad \quad  + \: EJ  \int_{0}^{L}   \alpha(s,t) \, k_{s}(s_{0}(t) + s) \, ds \,.
\end{aligned}
\label{formula}
\end{equation}
Equation \eqref{formula} shows that the dynamics of the rod is reduced to that of a point particle of mass $m$ subjected to a force given by the sum of  three terms.
The first one is a ``potential'' force depending exclusively on the geometry of $\Gam$, the second one is a friction term, while the third is an ``active'' force which depends on the spontaneous curvature $\alpha$. The following examples illustrate the role played by these terms in the dynamics of the system.

\subsection{Spiral channel}

\begin{figure}
\centering
	\includegraphics[width=0.90\textwidth]{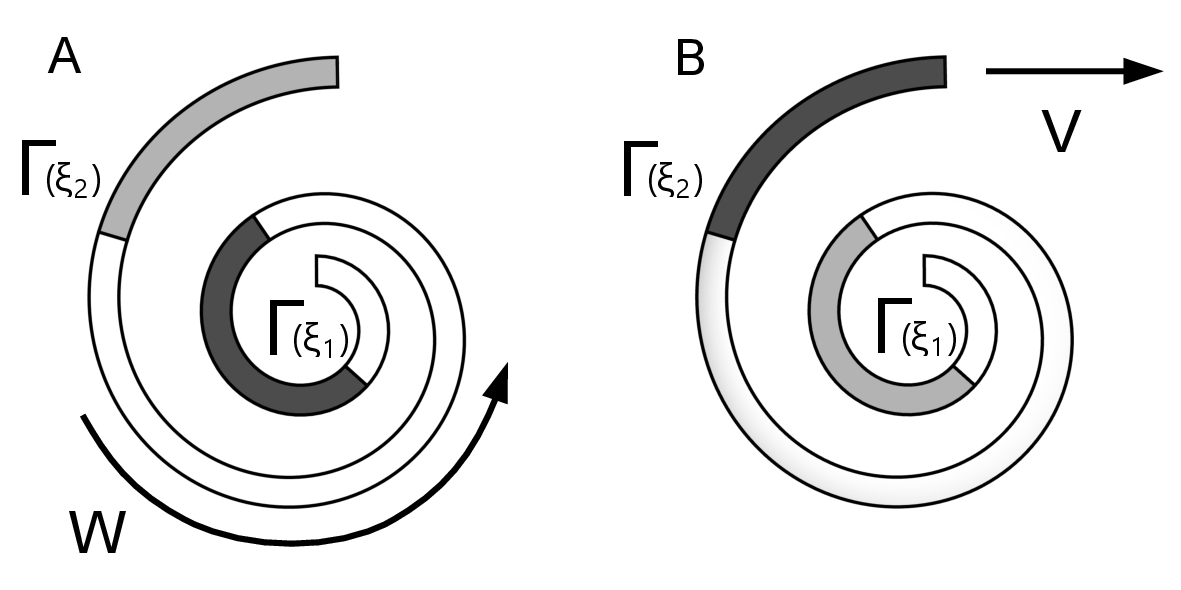}
	\caption{A) Two configurations of the elastic rod inside a spiral channel: initial (light grey) and final (dark grey). A positive work $W$ is necessary to vary the position of the end point from  $\Gam(\xi_{2})$ to  $\Gam(\xi_{1})$ and force the rod inside the channel. B) Upon release, the first end point slides back from  $\Gam(\xi_{1})$ to  $\Gam(\xi_{2})$ and the rod exits the channel with velocity $V$.}
	\label{2}
\end{figure}

Let us consider only the first term in the right hand side of \eqref{formula} by setting $\alpha , \gamma^{\pa}  =0$. The system described in this case is a passive elastic rod with straight rest configuration ($\alpha=0$) placed inside a curved channel with frictionless walls and no frictional interaction with the ground ($\gamma^{\pa}=0$). Observe that the only non-vanishing term in the right hand side of \eqref{formula} is the first one, which states that the driving force on the rod depends only on the curvature of the channel at the two ends of the body (this can be interpreted as a result of inextensibility).  Moreover, the sign of this force is such that the rod is always pushed towards the region of smaller curvature.  As an example,  consider the case of a spiral-shaped  channel where $k(s)=K/s$, with $K>0$  (see Fig.\ref{2}). Then (\ref{formula}) with $\alpha=0$ reads
\begin{displaymath}
	m \ddot{s}_{0}(t)=-U'(s_{0}(t)) \quad \textrm{where} \quad U(s_{0})= \frac{EJLK^{2}}{2(s_{0}+L)s_{0}} \, .
\end{displaymath}
In order to thread the rod inside the spiral by varying the coordinate of the end point from  $\xi_{2}$ to $\xi_{1}$ we have to do a positive work 
\begin{equation}
	W = U(\xi_{1})-U(\xi_{2}) = \frac{EJ}{2}  \left( \frac{LK^{2}}{(\xi_{1}+L)\xi_{1}}- \frac{LK^{2}}{(\xi_{2}+L)\xi_{2}} \right) > 0 \label{W}
\end{equation}
since we have to increase the curvature at every point of the body. If we then release the rod it will accelerate towards the exit and return back to $\xi_{2}$ with a positive velocity
\begin{equation}
	V = \sqrt{\frac{EJ}{m} \left( \frac{LK^{2}}{(\xi_{1}+L)\xi_{1}}- \frac{LK^{2}}{(\xi_{2}+L)\xi_{2}} \right)} > 0 \, . \label{V}
\end{equation}
The system moves towards a ``straighter'' configuration, decreasing its elastic energy and therefore increasing its kinetic energy. Similar problems of passive elastic rods sliding inside frictionless sleeves have been studied, both analytically and experimentally, in \cite{Bigoni}.

Let us suppose now that $\alpha , \gamma^{\pa} \neq 0$. The elastic rod can now vary its spontaneous curvature and it has to overcome a longitudinal frictional force to slide inside the spiral channel. The active force term in (\ref{formula}) can assume any value if we suppose that we have no restrictions in the choice of $\alpha$. Thus, in particular, an active elastic rod can slide \emph{inside} the spiral without need of external pushing. More generally, the system can achieve motion in a predetermined direction when placed inside any channel which does not present circular or straight sections of length greater than $L$. 
This last result is reminiscent of  theoretical and experimental findings of J. Gray in his study \cite{Gray} of snake undulatory locomotion. Using an energy balance argument, he concludes that it is possible for a snake to slide inside a channel closely fitting its body only provided such a channel exhibits a variation of curvature along its track. He then shows experimentally that snakes are able to move in sinusoidal closely fitting channels, but motion in straight ones only occurs through a different gait (concertina), which is impossible if the width of the channel and of the snake body are comparable.

We consider now two concrete examples of an active rod propelling itself inside the spiral channel. We do this by solving an inverse locomotion problem: we prescribe the motion of the rod and then deduce two histories of spontaneous curvatures that produce it. In this way, we also show non-uniqueness of the inverse problem.

\begin{figure}
\centering
	\includegraphics[width=0.90\textwidth]{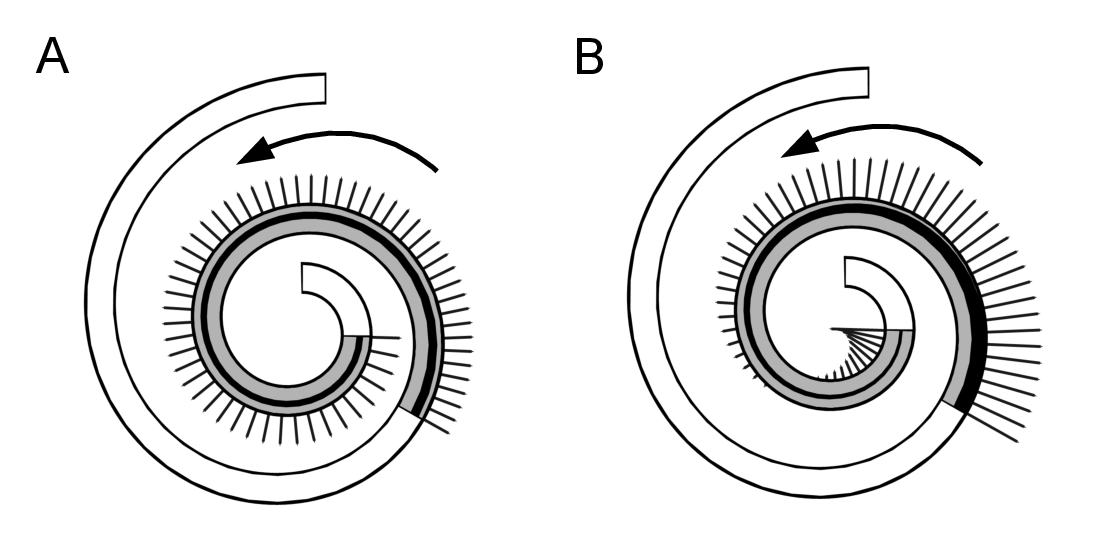}
	\caption{Snapshots for solutions generated by A) $\alpha_1$ and B) $\alpha_2$. Segments indicate the  magnitude of the transversal force exerted by the active rods on the channel. Spontaneous curvatures are represented through the shaded areas along the rod axis. Arrows indicate the direction of motion.}
	\label{3}
\end{figure}

Suppose we want to find an activation that propels the rod inside the spiral at constant velocity $\dot{s}_{0}(t)=-V<0$. Equation \eqref{formula} then reads
\begin{equation}
0 = \frac{EJ}{2}\left( \frac{K^2}{s_{0}(t)^2} - \frac{K^2}{(s_{0}(t)+L)^2}  \right) + \gamma^{\pa} L   -  EJ  \int_{0}^{L}   \frac{\alpha(s,t)K}{(s_{0}(t)+s)^2} \, ds 
\label{spir}
\end{equation} 
where $s_{0}(t)=s_{\textrm{in}} -V t$ for some initial value $s_{\textrm{in}}$ for $s_{0}$ at $t=0$. Set
\begin{gather*}
\alpha_{1}(s,t):= \frac{K}{s_{0}(t)+s} + \frac{\gamma^{\pa} (s_{0}(t)+s)^2}{EJ K} \quad \textrm{and} 
\\
\alpha_{2}(s,t):= \left( \frac{EJ}{2}\left( \frac{K^2}{s_{0}(t)^2} - \frac{K^2}{(s_{0}(t)+L)^2} \right)  + \gamma^{\pa} L \right) \frac{(s_{0}(t)+s)^2}{EJ K L}\, ,
\end{gather*}
then an easy calculation shows that \eqref{spir} is solved by both $\alpha=\alpha_1$ and $\alpha=\alpha_2$. Moreover, if the history of spontaneous curvatures $\alpha$ is given by  either $\alpha_{1}$ or $\alpha_{2}$, then $s_{0}(t)=s_{\textrm{in}} -V t$ becomes automatically a solution for the equations of motion \eqref{eqc1}-\eqref{eqc2}-\eqref{eqc3}, and $N$, $H$, and $f$ can be explicitly written following the procedure illustrated in the previous section. We denote by  $f_{1}$ and $f_{2}$ the lateral forces exerted by the active rod under the actuations $\alpha_{1}$ and $\alpha_{2}$, respectively. In Fig.\ref{3} the two solutions are illustrated with $L=0.5\textrm{m}$, $K=8\textrm{m}^{-1}$, $\gamma^{\pa}=0.3$, $EJ=10^{-3}\textrm{Nm}^{2}$, and where we set $\rho A V^2 k_{s}=0$, thereby ignoring inertial effects. An interesting consequence of this last assumption is that $f_{1}=\gamma^{\pa}(K + 2/K)$ is constant, as Fig.\ref{3}A shows. By contrast, $f_{2}$ 
 is not constant and it even changes sign (see Fig\ref{3}B).

Estimating lateral forces associated with internally actuated conformational changes, and minimizing them, may be of interest in the field of minimally invasive interventional medicine, e.g. for concentric-tube continuum robots, also called active cannulas \cite{Rucker}.

\subsection{Sinusoidal channel}

\begin{figure}
 \centering
	\includegraphics[width=1\textwidth]{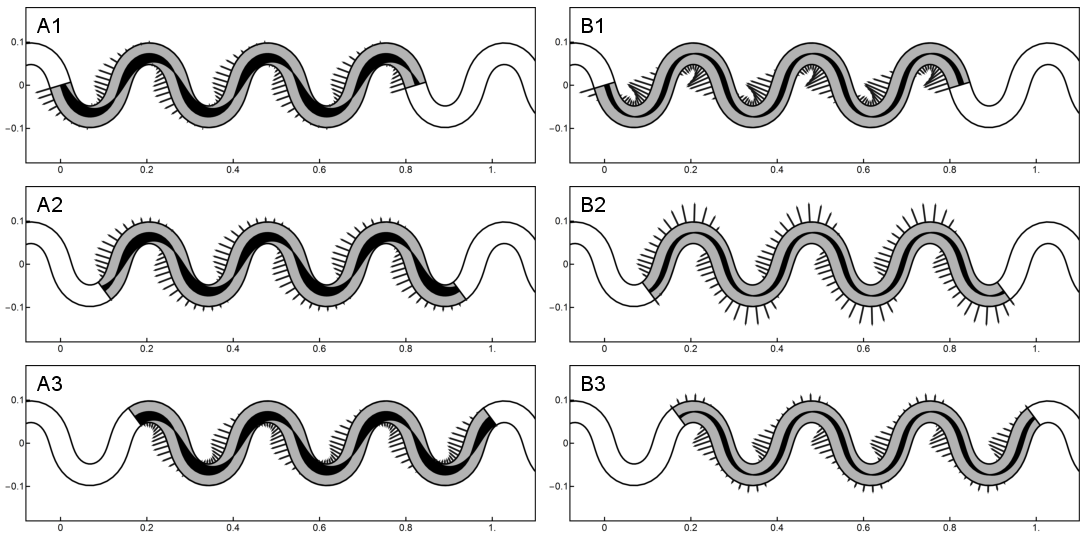}
	\caption{Snapshots for solutions generated by A) $\alpha_{\textrm{bend}}$ and B) $\alpha_{\textrm{act}}$  at three times: 1) $Vt/\lambda=0$, 2) $Vt/\lambda=2\pi/3$ and 3) $Vt/\lambda=4\pi/3$. Segments indicate the  magnitude of the transversal force exerted by the active rods on the channel. Spontaneous curvatures are represented through the shaded areas along the rod axis.}
	\label{4}
\end{figure}

We address in this section an inverse locomotion problem for a sinusoidal channel (see Fig.\ref{4}) meandering around the horizontal axis
\begin{equation}
	\Gam(\xi)= \int_{0}^{\xi}  \cos \Theta \E_{1} + \sin \Theta \E_{2} \: , \quad \textrm{where} \quad \Theta(\xi)= -\zeta \lambda \cos \left(\frac{\xi}{\lambda}\right) \label{path}
\end{equation}
and therefore
\begin{equation}
k(\xi)= \zeta \sin \left(\frac{\xi}{\lambda}\right)  \label{sin} \, .
\end{equation}
For small values of the geometric parameter  $\zeta$ the channel is close to a straight tube while, as  $\zeta$ grows, it becomes wavier and wavier. The wavelength $\lambda$ dictates how many turns the channel has per unit length.

We want to find  a history of spontaneous curvatures $\alpha(s,t)$ that produces motion along the sinusoidal channel \eqref{path} with constant longitudinal velocity
\begin{displaymath}
	\dot{s}_{0}(t) = V > 0 \, .
\end{displaymath}
Assuming that the trailing edge of the active rod lies at the origin at $t=0$, we must have $s_{0}(t)=Vt$. If we also assume that $L= 2\pi n \lambda$,  where $n$ is a positive integer, then the potential term in equation \eqref{formula} vanishes, and constant forward motion is realized only if the active force exactly matches the frictional one:
\begin{equation}
 \gamma^{\pa} L= EJ  \int_{0}^{L} \alpha(s,t)  k_s(s+Vt) ds= \frac{EJ \zeta}{\lambda} \int_{0}^{L} \alpha(s,t) \cos \left( \frac{s + V t}{\lambda} \right) \, ds  \, .
\label{alphachannel}
\end{equation}
We give two different solutions for the history of spontaneous curvatures $\alpha$ satisfying \eqref{alphachannel} (i.e. generating the same prescribed motion) by solving two constrained minimization problems. Among all $\alpha$'s such that \eqref{alphachannel}  holds, we find the ones that minimize $I_{\textrm{bend}}$ (the bending energy) and $I_{\textrm{act}}$ (the activity), where 
\begin{equation*}
I_{\textrm{act}}\left[\alpha\right] = \frac{1}{2} \int_{0}^{L} \! \! \alpha^{2}(s,t) \, ds \: \: \textrm{and} \: \:  I_{\textrm{bend}}\left[\alpha\right] =  \frac{EJ}{2} \int_{0}^{L} 
 \left( k(s + V t)
- \alpha(s,t) \right)^{2}  \! ds \, .
\end{equation*}
To solve, e.g., the second problem we consider the extended functional 
\begin{displaymath}
\hat{I}_{\textrm{bend}} \left[ \alpha ; q \right]:=	I_{\textrm{bend}}\left[\alpha\right] + q \int_{0}^{L} \alpha(s,t)  k_s(s+Vt)  \, ds 
\end{displaymath}
where $q$ is the Lagrange multiplier enforcing  \eqref{alphachannel}. The spontaneous curvature $\alpha_{\textrm{bend}}$ minimizing the bending energy $\hat{I}_{\textrm{bend}}$  must then solve $\delta\hat{I}_{\textrm{bend}} \left[ \alpha_{\textrm{bend}} ; q \right]=0$, where the variation of the extended functional is taken with respect to $\alpha$. A straightforward calculation gives
\begin{equation}
\alpha_{\textrm{bend}}(s,t)= \frac{q}{EJ}k_s(s+Vt) + k(s+Vt)\,,\quad q= \frac{\gamma^{\pa} L}{\int_0^L k^2_s(s+Vt) ds } \label{alphabend}
\end{equation}
where the second equality is obtained by plugging the expression for $\alpha_{\textrm{bend}}$ in \eqref{alphachannel}. More explicitly, using \eqref{sin}, we get
\begin{equation*}
	\alpha_{\textrm{bend}}(s,t) = \zeta \sin \left( \frac{s + V t}{\lambda} \right) +  \frac{q}{EJ} \cos \left( \frac{s + V t}{\lambda} \right) \quad \textrm{with} \quad 	q= \frac{\gamma^{\pa} L}{n \pi \zeta}  \, . 
\end{equation*}
From the equations of motion and the boundary conditions, taking again $\rho J=0$,  we then obtain
\begin{displaymath}
	H_{\textrm{bend}}=  - \frac{2 \gamma^{\pa} }{\zeta} \sin \left( \frac{s+ V t }{\lambda}\right) 
 \:, \quad
	N_{\textrm{bend}} = \frac{\gamma^{\pa} L}{4 \pi n} \Big(   \sin 2 \left( \frac{s+ V t }{\lambda}\right) + \sin 2 \, \frac{ V t }{\lambda} \, \Big)  \: ,
\end{displaymath}
and
\begin{align*}
f_{\textrm{bend}}(s,t) \: \: \: = & \: \: \: \gamma^{\pa} \frac{\zeta L}{4 \pi n} \Big(   \sin 2 \left( \frac{s+ V t }{\lambda}\right) + \sin 2 \, \frac{ V t }{\lambda} \, \Big) \sin \left( \frac{s+ V t }{\lambda}\right) \\
& \quad   - \gamma^{\pa} \frac{4 \pi n}{\zeta L}\cos\left( \frac{s+ V t }{\lambda}\right) - \rho A V^{2} \zeta \sin \left( \frac{s+ V t }{\lambda}\right) \, .
\end{align*}
Notice that none of the external and internal forces depend on the bending stiffness $EJ$. This allows us to consider the rigid limit $EJ \to \infty$, for which the observable motion and forces do not vary, while on the other hand $\alpha_{\textrm{bend}}(s,t) \to k(s +Vt)$. This limit case could be relevant for the steering of wheeled robots in which curvature control is achieved through internal motors.

Let us find $\alpha_{\textrm{act}}$ that minimizes $I_{\textrm{act}}$ by repeating the procedure above. We obtain in this case that the optimal $\alpha$ is proportional to the derivative of the channel's curvature $k_{s}$, whereby internal actuation is concentrated around inflexion points of the trajectory. This is reminiscent of patterns  of muscular activity observed in snake undulatory locomotion \cite{Jayne,Mo}. More in detail,
\begin{equation}
	\alpha_{\textrm{act}}(s,t) = \frac{q}{EJ} k_s(s+ V t )  = \frac{q}{EJ} \cos \left( \frac{s + V t}{\lambda} \right) \, , \label{alphaact}
\end{equation}
with $q$ given again by \eqref{alphabend}. In order to compare the two solutions we write
\begin{gather*}
	H_{\textrm{act}}(s,t) = H_{\textrm{bend}}(s,t) - EJ \, \frac{2 \pi n}{L} \cos \left( \frac{s+ V t }{\lambda}\right)\,,
\\
  N_{\textrm{act}}(s,t) = N_{\textrm{bend}}(s,t) + EJ \, \frac{\zeta^{2} \!}{4} \Big(   \cos 2 \left( \frac{s+ V t }{\lambda}\right) + \cos 2 \, \frac{ V t }{\lambda} \, \Big) \: ,
\end{gather*}
and
\begin{align*}
f_{\textrm{act}}(s,t) \: \: \: = & \: \: \: f_{\textrm{bend}}(s,t) + EJ \, \Big(\frac{2 \pi n}{L}\Big)^{2}  \zeta \sin \left( \frac{s+ V t }{\lambda}\right) \\  
& \quad  + EJ \, \frac{\zeta^{3} \!}{4} \Big(   \cos 2 \left( \frac{s+ V t }{\lambda}\right) + \cos 2 \, \frac{ V t }{\lambda} \, \Big) \sin \left( \frac{s+ V t }{\lambda}\right)  \end{align*}
for internal and external forces generated by $\alpha_{\textrm{act}}$, in terms of the corresponding quantities we found for  $\alpha_{\textrm{bend}}$. We observe that the 
two force fields differ by terms proportional to  $EJ$, while they become indistinguishable when $EJ\to 0$.

We give here a graphical representation of the two solutions, using material parameters taken from the zoological literature. Based on \cite{Mo} we set $L = 1.3\textrm{m}$ and $\gamma^{\pa}=\mu^{\pa} m g/L$, where $\mu^{\pa}= 0.2$ is the longitudinal friction coefficient, $m = 0.8\textrm{kg}$ and $g$ is the gravitational acceleration constant. Following \cite{HuShe}, we neglect the inertial terms in all the expressions setting $\rho A V^{2}=0$. As for the bending stiffness, we  explore a range going from  $EJ=10^{-4}\textrm{Nm}^{2}$  \cite{Long} to $EJ=10^{-3}\textrm{Nm}^{2}$  \cite{Mah}.

Results are shown in Fig.\ref{4}  for $n=3$ and $\zeta=18.5\textrm{m}^{-1}$, while we choose the largest value of $EJ$ in order to emphasise the difference between the two solutions in terms of forces exerted on the channel walls. The force field $f_{\textrm{bend}}$ consistently displays maxima in magnitude near the inflection points of curvature. On the other hand, $f_{\textrm{act}}$ varies substantially during motion: at some times it displays local maxima at the points of maximal concavity and convexity, while at some other times maxima are located at the inflection points. Notice that,  at points of maximal concavity and convexity of the rod, the lateral force $f$ is perpendicular to the horizontal axis (the average direction of motion) and does not contribute to propulsion.
We will comment further on these features in the next sections.

\section{The free-path case}

We now turn to the case in which the path is not a-priori known and study an active rod free to move on a flat surface through longitudinal sliding without lateral slipping. 
Accordingly, we impose the (non-holonomic) constraint
\begin{equation}
\R_{s}^{\bot} \cdot \R_{t} = 0
\label{nonhol}
\end{equation}
where $\R_{s}^{\bot} = \E_{3} \times \R_{s}$.  We denote by $- f \, \R_{s}^{\bot}$ the transversal reactive force per unit length (exerted by the ground on the rod) enforcing the no-slip condition, where $f$ is the Lagrange multiplier associated with constraint (\ref{nonhol}). At the same time, we suppose that a frictional force $\F^{\pa}$  given by \eqref{a02} acts in the longitudinal direction. 

This choice for $\F^{\pa}$ relies on the simplifying assumption that frictional forces are uniform along the rod's body. Moreover, real systems such as snakes \cite{Bau,Ber,HuShe} or snake-like robots \cite{Hi}, cannot rely on transversal frictional reactions of arbitrary magnitude  to prevent lateral slipping.  Solutions of interest for a more realistic description of these systems can be considered, for instance, those for which the reactive force $f$  imposing  constraint (\ref{nonhol}) does not exceed a maximum value, which can be determined experimentally.

\subsection{Derivation of the equations of motion}

We deduce the equations of motion through the Lagrange-d'Alembert principle, 
similarly to what is done in  \cite{GayB} and \cite{Va}. The principle states that, in the presence of the dissipative force $\F^{\! \! \Arrownot \, \, \Arrownot} \, \,$, a solution $(\R,\theta)$ that satisfies constraint (\ref{nonhol}) must solve
\begin{equation}
\delta  \int_{t_{1}}^{t_{2}}  \int_{0}^{L} \mathcal{L} \, ds dt + \int_{t_{1}}^{t_{2}}   \int_{0}^{L} \F^{\pa}  \cdot \delta \R  \, ds dt = 0
\label{LdA}
\end{equation}
for every variations $\delta \R$ and $\delta \theta$ that vanish at the boundary of $\left[0,L\right]\times\left[t_{1},t_{2}\right]$, while $\delta \R$ also satisfy
\begin{equation}
\R_{s}^{\bot} \cdot \delta \R = 0 \, .
\label{delta}
\end{equation}
Calculating the variation on the left hand side of (\ref{LdA}), after integration by parts and reordering, we have
\begin{align*}
& \delta  \int_{t_{1}}^{t_{2}}  \int_{0}^{L}   \mathcal{L} \, ds dt \: + \: \int_{t_{1}}^{t_{2}}  \int_{0}^{L} \F^{\pa}\cdot \delta \R(s,t) \, ds dt \: \:  = 
\\ 
& \quad \quad \quad \int_{t_{1}}^{t_{2}} \int_{0}^{L}  \! \left\{ - \rho A \, \R_{tt} +(\widetilde{N}\A)_{s}  +(\widetilde{H} \B)_{s} + \F^{\pa} \, \right\} \cdot \delta \R \, ds dt 
 \\
& \quad \quad \quad + \int_{t_{1}}^{t_{2}} \int_{0}^{L}    \left\{ - \rho J \theta_{tt} + EJ(\theta_{ss} - \alpha_{s}) + (\nu \widetilde{H} - \eta \widetilde{N}) \right\} \delta \theta \, ds dt \, .
\end{align*}
Since (\ref{LdA}) holds for all the variations satisfying (\ref{delta}), the coefficient multiplying $\delta \theta$ must vanish, while the coefficient relative to $\delta \R$ must take the form $f \R_{s}^{\bot}$, where  $f=f(s,t)$ is the unknown Lagrange multiplier enforcing constraint (\ref{nonhol}). The equations of motion then read
\begin{displaymath}
 \N_{s} +\F^{\pa} - f \R_{s}^{\bot} = \rho A \, \R_{tt}  \quad , \quad 
M_{s} \, \E_{3} + \R_{s} \times  \N = \rho J \theta_{tt} \E_{3} \, \: .
\end{displaymath}
We complement these equations with boundary conditions by relying again on the PoMBC.

\begin{figure}[h]
 \centering
		\includegraphics[width=1\textwidth]{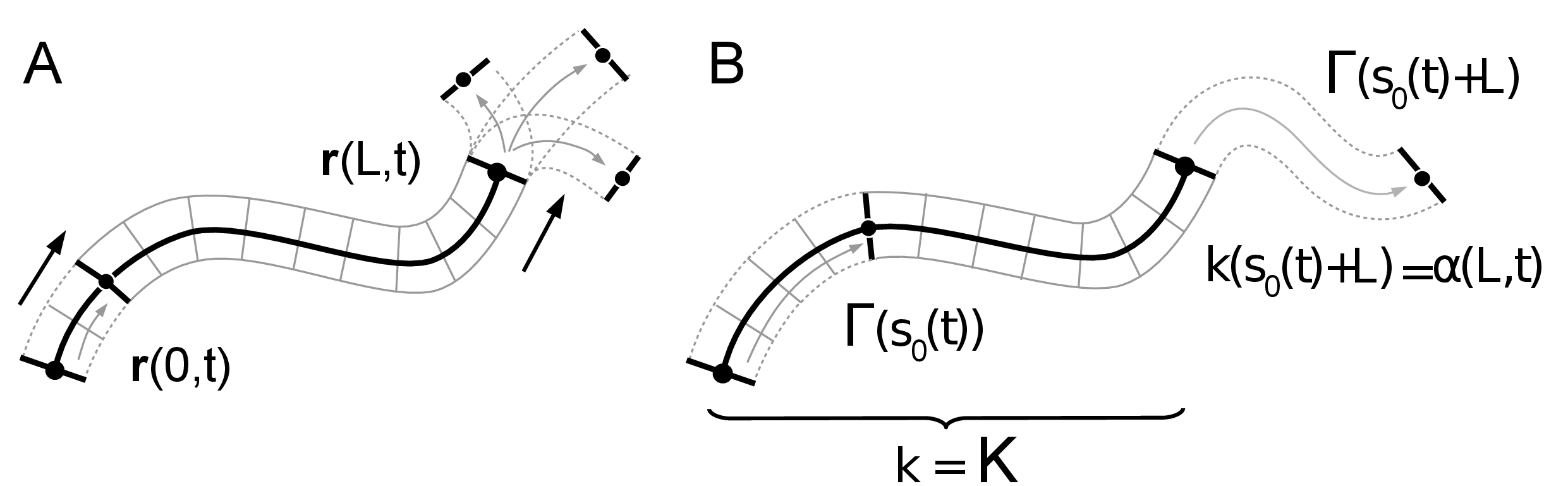}
	\caption{A) Sketch of the system moving while subjected to the constraint (\ref{nonhol}). Arrows indicate the direction of motion. Tail position   and  director change by assuming the values taken previously at an internal point. The head has an extra degree of freedom, since it is allowed to turn freely. B) Motion generated by a given spontaneous curvature history $\alpha(s,t)$. The curvature of the path at the leading edge is determined by the spontaneous curvature at $s=L$. Cross sections of the continuous rods are depicted through grey segments transversal to their midlines.}
\label{5} 
\end{figure}

Let us consider a typical configuration of the robot in motion while subjected to the external constraint (\ref{nonhol}). We suppose that such a movement is directed head-first, where we denote the head as $\R(L,t)$ and the tail as $\R(0,t)$. As shown in Fig.\ref{5}A, an asymmetry between head and tail emerges. Because of (\ref{nonhol}),  the tail position and  director can change only by assuming the values previously taken at an adjacent internal point. We can therefore impose on $s=0$ the same conditions we had in the channel case, namely, 
\begin{displaymath}
\R_{t} (0,t) = v_{0}(t) \, \R_{s}(0,t) \quad \textrm{and} \quad	\theta_{t} (0,t) =  v_{0}(t) k(0,t)
\end{displaymath}
where $v_{0}$ is the (only) generalized velocity at $s=0$ and $k(0,t)$ is the curvature of $\R$ evaluated in $s=0$ at time $t$. As for the head, since the path is no longer predetermined, we have an extra degree of freedom. Condition \eqref{nonhol} requires $\R_{t}$ and $\R_{s}$ to be collinear, therefore this extra degree of freedom must come from the rotation of the director. We then impose 
\begin{displaymath}
\R_{t} (L,t) = v_{L}(t) \, \R_{s}(L,t) \quad \textrm{and} \quad	
\theta_{t}(L,t) = \omega_{L}(t)
\end{displaymath}
where  $v_{L}$ and $\omega_{L}$ are the generalized velocities for the system at $s=L$. The work rate of the external edge forces is
\begin{equation*}
P_{\textrm{edge}}  =  \N(L,t) \cdot \R_{s}(L,t) v_{L}(t) + M(L,t) \omega_{L}(t) -	\Big( \N(0,t) \cdot \R_{s}(L,t) + M(0,t) k(0,t)\Big) \, v_{0}(t) \, . 
\end{equation*}
Thus, there are two generalized edge loads at $s=L$, namely, the axial tension $\N\cdot \R_{s}$ and the bending moment $M$, and one at $s=0$, with the same expression it had in the channel case. We  set the generalized loads equal to zero because, just like in the previous section, we suppose that no external edge force is doing work on the system.

Alongside with the boundary conditions coming from the vanishing of the generalized edge loads, the system must be complemented with equations (\ref{EB}) and (\ref{nonhol}).

The non-holonomic constraint (\ref{nonhol}) compels the active rod to move within a curve in the plane, much like it was for the channel case in the previous section. This time, however,  the path is not a-priori determined but is created during the motion, and it is an unknown of the problem. In fact, constraints (\ref{EB}) and (\ref{nonhol}) lead to the existence of some function $s_{0}$ and \emph{some} curve $\Gam$, which have \emph{both} to be determined, such that \eqref{RandTheta} holds. Since the  boundary conditions we derived hold only  for head-first motions, we only consider solutions satisfying 
\begin{equation}
\dot{s}_{0}(t) > 0 \, .
\label{headfirst}
\end{equation}
The equations of motion written in components are formally identical to the ones derived for the channel case
\begin{gather}
N_{s} - k H - \gamma^{\pa} = \rho A \, \ddot{s}_{0}(t) \label{eqs1}
\\
k N + H_{s} - f = \rho A \, k \, \dot{s}_{0}(t)^{2} \label{eqs2}
\\
 EJ(k_{s} - \alpha_{s})  + H = \rho J (k \, \ddot{s}_{0}(t) + k_{s} \, \dot{s}_{0}(t)^{2}  )  \label{eqs3}
\end{gather}
but  $k=k(s_{0}(t)+s)$ is no longer predetermined. On the other hand, the equations are closed through the boundary conditions obtained by setting the  three generalized edge loads equal to zero
\begin{equation}
\begin{aligned}
& N(0,t) + EJ\big(k(s_{0}(t)) - \alpha(0,t)\big)k(s_{0}(t) = 0 \, ,\\
& N(L,t)=0 \:  , \: \:   EJ\big(k(s_{0}(t)+L)-\alpha(L,t)\big)=0
\end{aligned}\label{BCs}
\end{equation}
together with the initial curvature $k(s_{0}(t_{0})+s)=K(s)$, with $s\in\left[0,L\right]$, and the initial values for $s_{0}$ and $\dot{s}_{0}$ at $t=t_{0}$. Such values must satisfy the compatibility relations
\begin{equation}
	\dot{s}_{0}(t_{0}) > 0 \: , \quad K(L)=\alpha(L,t_{0}) \: , \quad \textrm{and} \quad K_{s}(L)\dot{s}_{0}(t_{0}) =\dot{\alpha}(L,t_{0}) \, . \label{extra}
\end{equation}

In order to solve the locomotion problem, we need to find the unknown functions $N(s,t)$, $H(s,t)$, $f(s,t)$, together with $s_{0}(t)$ and $k(s_{0}(t)+s)$. The three equations of motion and the three boundary conditions \eqref{BCs} are sufficient to solve this problem uniquely. This leads to a unique solution also for $\R$ and $\theta$, once the initial position $\R(0,t_{0})$ and orientation $\theta(0,t_{0})$ of the first end are prescribed, by integrating the equations $\theta_s=k$ and $\R_s=\Gam$ as done, e.g.,  in \cite{HuShe,HuShe2}. The detailed proof is provided in the Appendix, and we only sketch here the heuristic argument behind it.

A key role is played by the third boundary condition in \eqref{BCs}, coming from the vanishing of the bending moment at the leading edge. This latter condition, namely,
\begin{equation}\label{steering_wheel}
 k(s_{0}(t)+L)=\alpha(L,t) 
\end{equation}
assigns a crucial role to the spontaneous curvature at the leading edge in determining the path followed by the system. Thus, the value of $\alpha$ at $s=L$ operates as a ``steering wheel'' while the internal values of the spontaneous curvature supply the active force for propulsion, as it was for the channel case.

Let us  see how $s_{0}$ and $k$ can be determined. There is no loss of generality if we take $t_{0}=0$ and $s_{0}(0)=0$. On the other hand, let us assume $\dot{s}_{0}(t)>0$ for $t \in \left[0,t^{*}\right)$ so that $s_{0}$ is invertible in the whole interval, and let's also assume that $t^{*}$ is small enough so that $s_{0}(t)<L$ for every $t$. Clearly,  $k(s)=K(s)$ is known for $s \in \left[0,L\right]$ from the initial condition. For $s>L$ we can recover $k$ from the history of spontaneous curvatures at the leading edge  because each point of the path $\Gam(\xi)$ with $\xi>L$ generated between $t_{0}$ and $t^{*}$ is the location of the leading edge at some time $s_{0}^{-1}(\xi-L)$, see Fig.\ref{5}B. Thus, setting
\begin{equation}\label{k}
	k(\xi) := \left\{ 
  \begin{array}{l l}
    K(\xi) & \quad \textrm{if $0\leq \xi \leq L$ }\\
    \alpha(L,s_{0}^{-1}(\xi-L)) & \quad \textrm{if $\xi \geq L$} 
  \end{array} \right .
\end{equation}
we can recover $k(s_{0}(t)+s)$ from the initial conditions, the given $\alpha$ and the knowledge of $s_{0}$. In turn, $s_{0}$ can be determined by substituting the expression for $H$ given by \eqref{eqs3} into \eqref{eqs1} and integrating with respect to $s$. Using \eqref{BCs}, we deduce
\begin{equation}
\begin{aligned}
\big(m + \rho J\, Q(s_{0}(t))\big) \, \ddot{s}_{0}(t) \: \:  = & \: \:  \frac{EJ}{2}\left( k^{2}(s_{0}(t)) - k^{2}(s_{0}(t) + L) \right)  - \gamma^{\pa} L \\
& \: \: - \rho J \, R(\dot{s}_{0}(t),s_{0}(t))  + EJ  \int_{0}^{L}    \alpha(s,t) \, k_{s}(s_{0}(t) + s) \, ds 
\end{aligned}\label{s0snake}
\end{equation}
where $R$ and $Q$ are given by \eqref{RandQ}. Moreover, using \eqref{k} and the change of variable $s = \xi - s_{0}(t)$, the last integral in \eqref{s0snake} can be written as the sum
\begin{displaymath}
	\int_{s_{0}(t)}^{L} \alpha(\xi - s_{0}(t),t) \, K_{s}(\xi) \, d \xi   + \int_{L}^{L +s_{0}(t)} \! \! \! \alpha(\xi - s_{0}(t),t) \, k_{s}(\xi) \, d \xi \, .
\end{displaymath}
The second summand in the last expression can be rewritten further, using the change of variable $\xi=s_{0}(\tau)+L$, as
\begin{align*}
\int_{L}^{L +s_{0}(t)}   \alpha(\xi - s_{0}(t),t) \, k_{s}(\xi) \, d \xi & \: \: =  \: \:   \int_{0}^{t}  \alpha(s_{0}(\tau) - s_{0}(t) + L,t) \, k_{s}(s_{0}(\tau) + L )\dot{s}_{0}(\tau) \, d \tau \\
& \: \: =  \: \:  \int_{0}^{t} \! \alpha(s_{0}(\tau) - s_{0}(t) + L,t) \, \dot{\alpha}(L,\tau) \, d \tau 
\end{align*}
where we have used the identity $k_{s}(s_{0}(t) + L )\dot{s}_{0}(t) =\dot{\alpha}(L,t)$ following from \eqref{steering_wheel}. Finally, observing that in view of our assumption $s_{0}(t)<L$, we have $k(s_{0}(t))=K(s_{0}(t))$ and $k(s_{0}(t)+L)=\alpha(L,t)$, it follows that
\begin{equation*}
	R   =  \frac{\dot{s}_{0}(t)^{2}}{2} \left(\alpha^{2}(L,t) - K^{2}(s_{0}(t)) \right)  \: \: \:  \textrm{and} \: \: \:
	Q  = \int_{s_{0}(t)}^{L} \! \! K^{2}(\xi) d\xi + \int_{0}^{t} \alpha^{2}(L,\tau)\dot{s}_{0}(\tau)\, d \tau \, .
\end{equation*}
Equation \eqref{s0snake} is in fact
\begin{align*}
\big(m + \rho J\, Q\big) \, \ddot{s}_{0}(t)  \: = & \: \frac{EJ}{2}\left( K^{2}(s_{0}(t)) - \alpha^{2}(L,t) \right)  - \gamma^{\pa} L  \nonumber \\
 & \: - \rho J \, R \nonumber + EJ \int_{s_{0}(t)}^{L} \alpha(\xi - s_{0}(t),t) \, K_{s}(\xi) \, d \xi  \\ 
 & \: +  EJ \int_{0}^{t} \alpha(s_{0}(\tau) - s_{0}(t) + L,t) \, \dot{\alpha}(L,\tau) \, d \tau \nonumber
\end{align*}
an integro-differential equation in $s_{0}$ alone which can be uniquely solved in terms of the data of the problem, as proven in the Appendix. 

Just like in the channel case, once $s_{0}$ and $k$ are known, the unknown functions $H$, $N$ and $f$ can be readily deduced from \eqref{eqs3}, \eqref{eqs1}, and \eqref{eqs2} respectively.

\subsection{Serpentine solutions}

In this section we provide a class of explicit serpentine solutions for the free-path locomotion problem, by exploiting solutions constructed for the channel case. We obtain these solutions by solving an inverse locomotion problem, prescribing the motion first and then looking for a history of spontaneous curvatures $\alpha(s,t)$ that produces it. 

Let us consider the sinusoidal path $\Gamma$ given by \eqref{path} and assume that our active rod slides at constant longitudinal velocity $V$, so that $s_{0}(t)=Vt$. As we did before, we set $\rho J=0$ for simplicity. Following the arguments of Section 3(c) we conclude that $\alpha$ must again solve \eqref{alphachannel}. In addition, we must now also require the boundary condition \eqref{steering_wheel}, which assigns the steering role to $\alpha$, to be satisfied. Notice that none of the spontaneous curvatures we obtained in the channel case fulfils \eqref{steering_wheel}. However, as we show in the following,  we can locally modify any $\alpha$ solving \eqref{alphachannel} so that  \eqref{steering_wheel}  is also satisfied. 

We focus below on the history of spontaneous curvatures $\alpha_{\textrm{act}}$ given by \eqref{alphaact} since it is the one that more closely resembles the typical muscular activity patterns observed in undulating snakes. If we consider a function
\begin{equation}
\alpha(s,t) = \alpha_{\textrm{act}}(s,t) + \widetilde{\alpha}(s,t)
\label{alphasnake}
\end{equation}
with a ``steering term''  $\widetilde{\alpha}$ such that
\begin{equation}\label{alphasteer}
 \widetilde{\alpha}(L,t) = \zeta \sin \left( \frac{L+ V t}{\lambda} \right) - \alpha_{\textrm{act}}(L,t) \quad \textrm{and} \quad  \int_{0}^{L} \widetilde{\alpha}(s,t) \cos \left( \frac{s + V t}{\lambda} \right)\, ds =0 \, ,
\end{equation}
then $\alpha$ satisfies both \eqref{alphachannel} and \eqref{steering_wheel}.  With $\alpha$ having these properties, $s_{0}(t)=Vt$ becomes a solution for the equations of motion, and the expression for $N$, $H$ and $f$ can be deduced following the procedure of Section 3(a). 

The  term  $\widetilde{\alpha}$ in \eqref{alphasnake} satisfying \eqref{alphasteer} can be taken of the form
\begin{equation}\label{steer}
	\widetilde{\alpha}(s,t) =  \Bigg\{ \begin{array}{l}
		0 \: \: \textrm{if $s \in \left[0,L-\delta \right]$} \\
	\sum_{i=3}^{Q}  p_{i}(t)(s-L+\delta )^{i}  \: \:	\textrm{if $s \in \left[L-\delta ,L \right]$}
	\end{array} 	
\end{equation}
where $\delta$ is an arbitrary constant, which can be set as small as we want, and $p_{i}(t)$ with $i=3,\ldots,Q$ are coefficients explicitly depending on $t$ and implicitly depending also on $\delta$ and all the dynamical parameters. These coefficients can be uniquely determined imposing \eqref{alphasteer} and any other $Q-5$ linearly independent relations between them (for example, in the numerical experiment we are about to propose, we imposed $\widetilde{\alpha}_{ss}(L,t)=0$, which led to a smooth generated force field $f$ concentrated near the head). Notice that the function $\widetilde{\alpha}$ so defined is twice continuously differentiable. 

If we take $\delta$ small enough, then $\alpha$ differs from $\alpha_{\textrm{act}}$ only in a small neighbourhood of the leading edge where the steering term $\widetilde{\alpha}$ is non zero. The reactive shear force and tension are now given by
\begin{gather*}
	H(s,t)= H_{\textrm{act}}(s,t) + EJ\widetilde{\alpha}_{s}(s,t) \: \: \textrm{and}  
\\
	N(s,t) = N_{\textrm{act}}(s,t)	+  EJ \zeta \int_{0}^{s} \!  \sin \left( \frac{\xi + V t}{\lambda} \right)\widetilde{\alpha}_{s}(\xi,t) \, d\xi 
\end{gather*}
while the force exerted on the ground reads, in this case,
\begin{equation*}
f(s,t) =  f_{\textrm{act}}(s,t) + EJ \zeta^{2} \int_{0}^{s}   \sin \! \left( \frac{\xi+ V t }{\lambda}\right) \widetilde{\alpha}_{s}(\xi,t) \, d\xi \sin \left( \frac{s+ V t}{\lambda} \right) + EJ\widetilde{\alpha}_{ss}(s,t) \, . 
\end{equation*}
From the last equalities it follows that, if $\delta$ is small, forces (external and internal) have  the same values of the corresponding ones obtained in the channel case with the exception of a small region near the leading edge.

\begin{figure}[!h]
 \centering
	\includegraphics[width=1\textwidth]{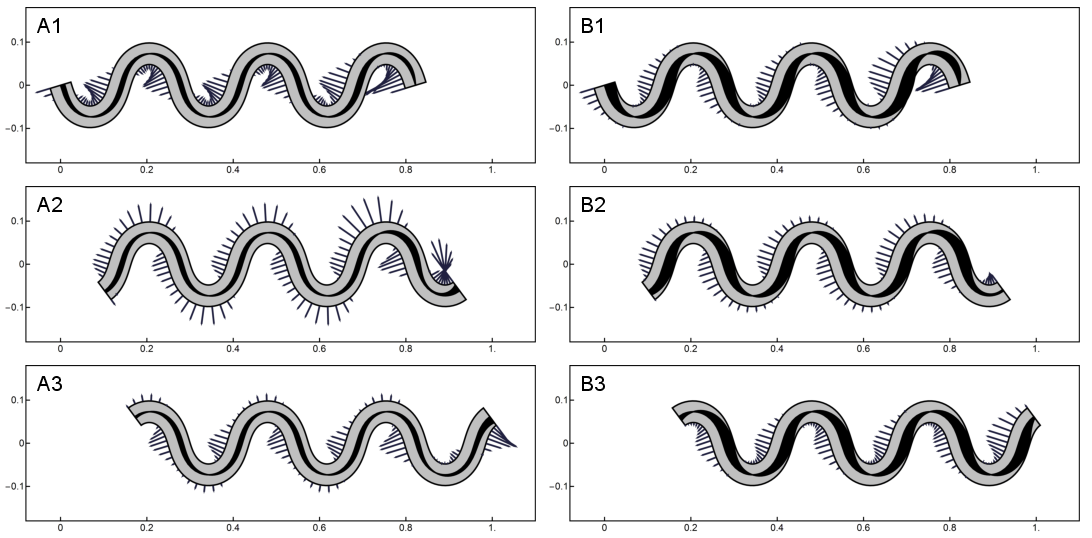}
	\caption{Solutions with different bending stiffness, A) $EJ=10^{-3}\textrm{Nm}^{2}$ and B) $EJ=10^{-4}\textrm{Nm}^{2}$, at three times: 1) $Vt/\lambda=0$, 2) $Vt/\lambda=2\pi/3$ and 3) $Vt/\lambda=4\pi/3$. Segments indicate the  magnitude of the transversal force exerted on the ground surface. Spontaneous curvatures are represented through the shaded areas along the rod axis. To help visualization, the spontaneous curvatures are here not drawn to scale: the maximal width of the dark shades in B) should be ten times greater than A).}
	\label{6}
\end{figure}

\begin{figure}[!h]
 \centering
	\includegraphics[width=1\textwidth]{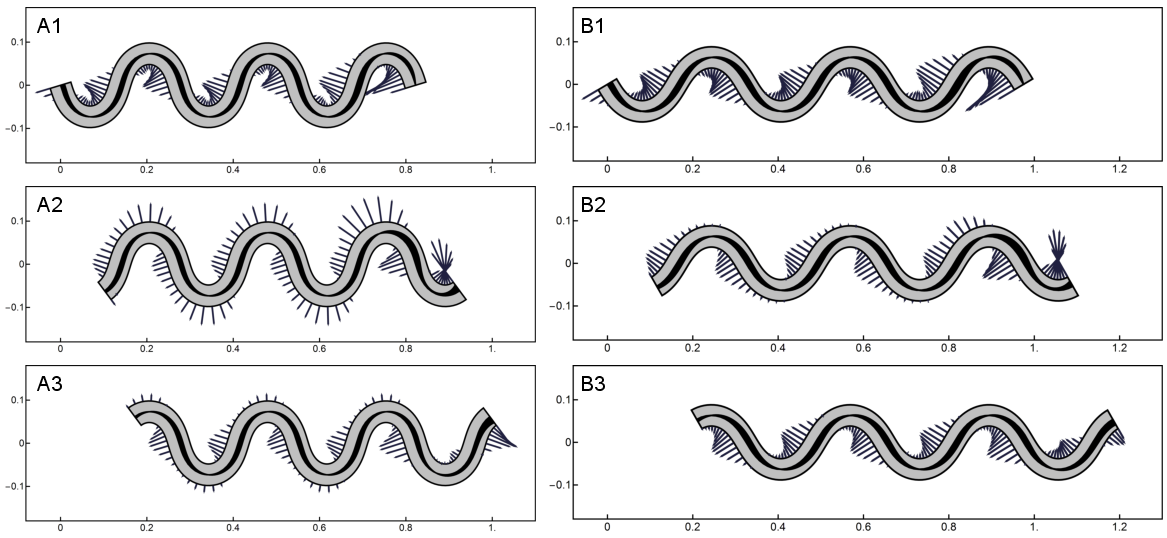}
	\caption{Solutions with the same bending stiffness and different path geometries, A) $\zeta=18.5\textrm{m}^{-1}$ and B) $\zeta=15\textrm{m}^{-1}$, at three times: 1) $Vt/\lambda=0$, 2) $Vt/\lambda=2\pi/3$ and 3) $Vt/\lambda=4\pi/3$. Segments indicate the  magnitude of the transversal force exerted on the ground surface. Spontaneous curvatures are represented through the shaded areas along the rod axis.}
	\label{7}
\end{figure}

We set $\delta/L=0.25$ and we give here two graphical comparisons of the same solution fitted with different parameters (Fig.\ref{6} and Fig.\ref{7}). 

In Fig.\ref{6}A and Fig.\ref{7}A we take the same values we considered in Section 3(c) for all the parameters. When compared with that of Fig.\ref{4}B,  this solution clearly shows the asymmetry that the steering term $\widetilde{\alpha}$ generates in the activation and force pattens in the proximity of the head (leading edge). In Fig.\ref{6}B we take the smaller value for the bending stiffness, $EJ=10^{-4}\textrm{Nm}^{2}$. Notice that this solution displays a similar force pattern to that of Fig.\ref{4}A (as expected from the formulas we derived in Section 3(c)), which is generated by a different spontaneous curvature history. Finally, in Fig.\ref{7}B, we consider an active rod with the same bending stiffness but moving with a less tortuous gait (smaller $\zeta$). Observe that, also in this case, we obtain an almost stationary force pattern, which is qualitatively similar to that of Fig.\ref{6}B.

Summarizing, we see a picture consistent with that of snake undulatory locomotion hypothesized in \cite{Mah} (muscular activity and  lateral forces both concentrated  near the inflection points of the trajectory, where the propulsive effect of the lateral forces is largest because their component along the  average direction of motion is largest) emerges either automatically,  for specific choices of  material parameters (Fig.\ref{6}B), or through adjustment of the gait (Fig.\ref{7}B). Lateral forces near points of maximal and minimal convexity may also be ruled out by eliminating ground contact (by lifting  portions of the body near those points), as it is done in \cite{HuShe,HuShe2} and sometimes observed in undulating snakes.

\section{Discussion}

We have studied the motion of an active rod  (a planar inextensible elastic rod of finite length with adjustable spontaneous curvature)
arising from the interaction between external constraints and internal actuation by spontaneous curvature. 
Using Cosserat theory, we have formulated and solved both direct and inverse locomotion problems for two cases: one in which the system is forced to move along a prescribed path, and the other in which the path is not fixed a-priori and the system slides along its tangential direction while subjected to lateral forces preventing lateral slipping. 
We have obtained a procedure to generate free-path solutions from solutions with prescribed-path, by recognising the dual role (pushing and steering) played by spontaneous curvature in powering undulatory locomotion of the rod. Finally, we have obtained explicit analytic solutions and  formulas that can be used to study the connections between observed motion, internal actuation, and forces transmitted to the environment, and to explore how these connections are affected by the mechanical properties of the system (its bending stiffness).

Although our results hold for a (very specific) model system, 
it may be interesting to compare some of them with observations made in the context of undulatory locomotion of snakes.
For this exercise to make  sense, we are formulating the implicit assumption that our mechanism of internal actuation by spontaneous curvature can provide a reasonable proxy for muscular actuation, and that the free-path motion of the organism we are considering does not cause lateral slipping, but only involves longitudinal sliding (as it is sometimes observed).

The first example is formula \eqref{formula}, which provides a compact summary of some classical observations on snake locomotion by Gray \cite{Gray,GrayLiss}. Undulatory locomotion in closely fitting channels is possible only if the channel presents a variation of curvature along its track. The formula
 explains the mechanism by which spontaneous curvature can provide the driving force for locomotion inside a tightly fitting channel, and our analysis delivers formulas to calculate the lateral forces exerted on the channel walls. It would be interesting to compare these with experimental measurements.

A second example is the observation that, among various possible actuation strategies producing the same prescribed motion, the one minimising actuation effort (as measured by the integral norm of spontaneous curvature) is proportional to the arc-length derivative of the curvature of the trajectory. This means that local actuation is maximal at the inflexion points of the trajectory, and zero at points of maximal and minimal curvature. This is closely reminiscent of the typical pattern of muscular actuation emerging from experimental measurements on snakes \cite{Jayne,Mo}, and it would be interesting to explore further  the  reasons behind this analogy.

Finally, our analysis suggests that the connection between observed motions, internal actuation, and transmitted forces may be strongly affected by the passive mechanical properties of the system, such as its bending stiffness. The conceptual picture of snake undulatory locomotion in which both muscular activity and  lateral forces are concentrated  near the inflection points of the trajectory, previously theorised in \cite{Mah}, can emerge either automatically,  for specific choices of  material parameters, or through the adjustment of the gait

Understanding the mechanisms that control gait selection and, in particular, whether  there are optimality criteria explaining it in biological organisms, and whether some of them may be useful for the engineering of artificial devices represent interesting challenges for future work (see however \cite{HuShe2,Mah} for results in this direction). Adding some important ingredients, currently not present in our model, may prove necessary. One example is some form of
active local control of the frictional interactions between body and ground, as is done in \cite{HuShe,HuShe2}.
Moreover, when considering real snakes behaviour it is natural to speculate  that muscular activity may be, at least to some extent,  a reaction to external stimuli (the forces exerted by the ground on the snake), thereby creating an interplay between the two dynamical variables. It would be interesting to study how our model could be extended to account for such  feedback mechanisms. 
All these questions will require further study.

\section*{Acknowledgements}
This study was started after an inspiring lecture given by prof. D. Bigoni at the `Material Theories' workshop held in 2013 at the Mathematisches Forschungsinstitut Oberwolfach. We thank prof. F. L. Chernousko for pointing out to us important references in the Russian literature. ADS thanks  F. Renda for useful discussions. This work was supported by the European Research Council through the ERC Advanced Grant 340685-MicroMotility.

\newpage

\section*{Appendix. Existence and uniqueness of free-path solutions}
For the sake of simplicity we take $\rho J=0$. The following arguments can be easily adjusted for the general case.

Suppose we have a solution of the free-path locomotion problem \eqref{eqs1}-\eqref{eqs2}-\eqref{eqs3} satisfying the boundary conditions (\ref{BCs}) and the extra requirements (\ref{headfirst}) and (\ref{extra}) where $K(s)=k(s_{0}(t_{0})+s)$ for $s\in\left[0,L\right]$. Again, there is no loss of generality taking $t_{0}=0$ and $s_{0}(t_{0})=0$. Let us first suppose, by restricting its domain of definition if necessary, that $s_{0}$ is defined on a time interval $\left[0,t^{*}\right)$ such that $s_{0}(t)\leq L$ for every $t \in \left[0,t^{*}\right)$.
The arguments  used in Section 4(a) show that, within this time interval, the equation for $s_{0}$ in terms of the data of the problem reads
\begin{equation}
\begin{aligned}
m \ddot{s}_{0}(t) \: = & \: \frac{EJ}{2}\left( K^{2}(s_{0}(t)) - \alpha^{2}(L,t) \right)  - \gamma^{\pa} L  + EJ \int_{s_{0}(t)}^{L} \alpha(\xi - s_{0}(t),t) \, K_{s}(\xi) \, d \xi  \\
                     & \quad   +  EJ \int_{0}^{t} \alpha(s_{0}(\tau) - s_{0}(t) + L,t) \, \dot{\alpha}(L,\tau) \, d \tau \, .
\end{aligned}  \label{s0app}
\end{equation}
The general case in which $s_{0}(t)> L$ may also occur (i.e., the trailing edge is no longer contained in the image of the initial configuration, see Fig.\ref{5}B)  can be handled by applying the following simple, yet technical, argument. As we will show, a local solution  $s_{0}$  for \eqref{s0app} exists and is unique once a positive initial velocity $\dot{s}_{0}(t_{0})$ is given, by requiring that $s_{0}(t)\leq L$. If a local solution $s_{0}$ of \eqref{s0app} exists and is unique then, for every given initial conditions, there exists only one solution with a maximal interval of definition. For such maximal solutions we can have either of  two cases. In the first, the maximal interval of existence of $s_{0}$ with $\dot{s}_{0}>0$ is of the type $\left[0,t^{*}\right)$ and, for every $t$ in the interval, $s_{0}(t)\leq L$  holds. If that occurs, the only solution of the free-path problem is defined in the time interval $\left[0,t^{*}\right)$, the curvature $k$ can be derived through \eqref{k} while all the other unknowns can be deduced by the the same procedure we employed in the channel case in Section 3(a). In the second case, a solution $s_{0}$ of \eqref{s0app} satisfying $\dot{s}_{0}>0$ and $s_{0}(t)\leq L$  can be only defined in a maximal domain of the type $\left[0,t^{*}\right]$. For a solution of this kind we must have $s_{0}(t^{*})=L$, by maximality. In this last case we can still define $k$ through \eqref{k} as we did before. Then we can reapply all the arguments of Section 4(a) finding an equation of the type \eqref{s0app} for a new variable $s_{0}^{*}$ with new initial conditions for the free-path locomotion problem, namely $s_{0}^{*}(t^{*})=s_{0}(t^{*})$, $\dot{s}_{0}^{*}(t^{*})=\dot{s}_{0}(t^{*})$ and the new initial curvature $K^{*}(s)=k(L+s)$  with $s\in\left[0,L\right]$. After that we are able to solve the new integro-differential problem uniquely for $s_{0}^{*}$, recover the value for all the unknowns, and then glue the solutions together. We re
 peat this procedure until we reach eventually a maximal domain of existence for the general solution.

The existence and uniqueness of free-path locomotion solutions then follows from the local existence and uniqueness of solutions of \eqref{s0app} with the extra requirements $\dot{s}_{0}(t)>0$ and $s_{0}(t)\leq L$. This can be proved using standard contraction mapping arguments. The result holds under the very reasonable assumption of $\alpha$ and $K$ being differentiable and uniformly bounded together with their derivatives. 

Observe that we can recast \eqref{s0app} into a set of integro-differential equations of the form
\begin{equation}
	\dot{\X}(t) = \G (\X(t),t) + \int_{0}^{t} \h (\X(\tau)-\X(t),\tau,t) \, d \tau \, \label{integro},
\end{equation}
with 
\begin{align*}
   & \X(t)=(x(t),y(t)) \: \:  , \: \: \h(\X,\tau,t)  = \Big( \: 0 \: , \:  EJ \alpha(x + L,t) \, \dot{\alpha}(L,\tau)   \: \Big)   \quad \textrm{and} \\
	  \G (\X,t)  = & \left( \: \frac{y}{m} \:  , \: \frac{EJ}{2}\left( K^{2}(x) - \alpha^{2}(L,t) \right)   - \gamma^{\pa} L  +  EJ \int_{x}^{L} \alpha(\xi - x,t) \, K_{s}(\xi) \, d \xi   \: \right) \, . 
\end{align*}
It is clear that $\X$ solves \eqref{integro} if and only if $s_{0}(t):=x(t)$ solves \eqref{s0app}. We first extend $K$ and $\alpha$ outside $\left[0,L\right]\times\left[0,\infty\right)$ while keeping their regularity properties. Then we consider the Cauchy problem for (\ref{integro}) with initial conditions $\X(0)=\X_{0}$ and no extra assumption on the solutions besides differentiability. The problem can be easily proved to be equivalent to  that of the existence of a fixed-point for the operator $C$ defined as
\begin{displaymath}
	C\left[\X\right](t) = \X_{0} + \int_{0}^{t} \left[ \G (\X(\lambda),\lambda) + \int_{0}^{\lambda} \h (\X(\tau)-\X(\lambda),\tau,\lambda) \, d \tau  \right] \, d\lambda  \, .
\end{displaymath}
We restrict the operator $C$ to the space $B_{\X_{0}}^{M,T}$ of continuous vector valued functions $t \mapsto \X(t)$ defined on the interval $t\in\left[0,T\right]$ and such that 
\begin{displaymath}
	\left\|\X-\X_{0}\right\|=\max_{t \in \left[0,T\right]}\left|\X(t)-\X_{0}\right| \leq M     \, . 
\end{displaymath}
There is no loss of generality in assuming that the extensions we considered for $K$ and $\alpha$ lead to the existence of two constants $M_{\G}$ and $M_{\h}$ such that
\begin{displaymath}
	\left|\G(\X(t),t)\right| \leq M_{\G} \quad \textrm{and} \quad \left|\h (\X(\tau)-\X(t),\tau,t) \right| \leq  M_{\h} \quad  \textrm{for every $\tau$ and $t$}
\end{displaymath}
and for every $\X \in B_{\X_{0}}^{M,T}$.  We can also assume that there are other two constants $L_{\G}$ and $L_{\h}$ such that
\begin{displaymath}
	\left|\G(\X,t) - \G(\Y,t) \right| \leq L_{\G}|\X -\Y | \quad \textrm{and}  \quad \left|\h (\X,\tau,t) - \h(\Y,\tau,t)\right| \leq  L_{\h}|\X -\Y |
\end{displaymath}
for every $\tau$ and $t$ and for every $\X , \Y \in B_{\X_{0}}^{M,T}$. We have then
\begin{displaymath}
\left|	\, C\left[\X\right](t) - \X_{0} \right| \leq T M_{\G} + T^{2} M_{\h}
\end{displaymath}
and also
\begin{eqnarray*}
\left|	\, C\left[\X\right](t) -  C\left[\Y\right](t)  \right| & \leq &  \left| \int_{0}^{t}  \G (\X(\lambda),\lambda) -  \G (\Y(\lambda),\lambda) \, d\lambda \, \right| \\
 &   &  + \: \: \left| \int_{0}^{t}    \int_{0}^{\lambda} \h (\X(\tau)-\X(\lambda),\tau,\lambda)-\h (\Y(\tau)-\Y(\lambda),\tau,\lambda) \, d \tau   \, d\lambda  \, \right| \\
& &\leq \: \: T L_{\G} \left\|\X - \Y \right\| + L_{\h}  \int_{0}^{t}    \int_{0}^{\lambda} \left| \X(\tau)-\X(\lambda) - \big( \Y(\tau)-\Y(\lambda) \big) \right| \, d \tau   \, d\lambda \\
&  & \\
&  & \leq \: \: (T L_{\G} +2T^{2}L_{\h} )\left\|\X - \Y \right\| \, .
\end{eqnarray*} 
For small enough $T$  the operator $C$ is a contraction from $B_{\X_{0}}^{M,T}$ into itself, therefore it has only one fixed point. This proves local existence and uniqueness for the extended version of (\ref{integro}). If we take $\X_{0}=(0,y_{0})$ with $y_{0}>0$ then, restricting the domain of existence to an interval $\left[0,T^{*}\right)$ if necessary,  we have by continuity  $x(t)\leq L$ and $\dot{x}(t)=y(t)>0$ for every $t\in\left[0,T^{*}\right)$, hence obtaining  the unique solution to the original  (non-extended) problem.

\end{document}